\documentstyle[epsf,eqsecnum,floats,preprint,aps]{revtex}

\tighten

\input epsf
\begin{document}

\newcommand{\be}{\begin{equation}}
\newcommand{\ee}{\end{equation}}
\newcommand{\bea}{\begin{eqnarray}}
\newcommand{\eea}{\end{eqnarray}}
\newcommand{\PSbox}[3]{\mbox{\rule{0in}{#3}\includegraphics{#1}\hspace{#2}}}

\def\5M{M^3_{(5)}}
\def\4M{M^2_{(4)}}

\overfullrule=0pt
\def\Int{\int_{r_H}^\infty}
\def\d{\partial}
\def\e{\epsilon}
\def\M{{\cal M}}
\def\high{\vphantom{\Biggl(}\displaystyle}
\catcode`@=11
\def\@versim#1#2{\lower.7\p@\vbox{\baselineskip\z@skip\lineskip-.5\p@
    \ialign{$\m@th#1\hfil##\hfil$\crcr#2\crcr\sim\crcr}}}
\def\simge{\mathrel{\mathpalette\@versim>}} %
\def\simle{\mathrel{\mathpalette\@versim<}} %
\catcode`@=12 

\rightline{NYU-TH 01/11/01}
\rightline{hep-th/0111168}
\vskip 3cm

\setcounter{footnote}{0}

\begin{center}
\large{\bf Cosmic Strings in a Braneworld Theory with Metastable Gravitons}
\ \\
\ \\
\normalsize{Arthur Lue\footnotetext{E-mail:  lue@physics.nyu.edu}}
\ \\
\ \\
\small{\em Department of Physics \\
New York University \\
New York, NY 10003}

\end{center}

\begin{abstract}

\noindent
If the graviton possesses an arbitrarily small (but nonvanishing)
mass, perturbation theory implies that cosmic strings have a nonzero
Newtonian potential.  Nevertheless in Einstein gravity, where the
graviton is strictly massless, the Newtonian potential of a cosmic
string vanishes.  This discrepancy is an example of the
van~Dam--Veltman--Zakharov (VDVZ) discontinuity.  We present a
solution for the metric around a cosmic string in a braneworld theory
with a graviton metastable on the brane.  This theory possesses those
features that yield a VDVZ discontinuity in massive gravity, but
nevertheless is generally covariant and classically self-consistent.
Although the cosmic string in this theory supports a nontrivial
Newtonian potential far from the source, one can recover the Einstein
solution in a region near the cosmic string.  That latter region grows
as the graviton's effective linewidth vanishes (analogous to a vanishing
graviton mass), suggesting the lack of a VDVZ discontinuity in this
theory.  Moreover, the presence of scale dependent structure in the
metric may have consequences for the search for cosmic strings through
gravitational lensing techniques.
\end{abstract}

\setcounter{page}{0}
\thispagestyle{empty}
\maketitle

\eject

\vfill

\baselineskip 18pt plus 2pt minus 2pt

\section{Introduction}
\setcounter{footnote}{0}

General relativity is a theory of gravitation that supports a massless
graviton with two degrees of freedom.  However, if one were to
describe gravity with a massive tensor field, general covariance is
lost and the graviton would possess five degrees of freedom.  In the
limit of vanishing mass, these five degrees of freedom may be
decomposed into a massless tensor (the graviton), a massless vector (a
graviphoton which decouples from any conserved matter source) and a
massless scalar.  This massless scalar persists as an extra degree of
freedom in all regimes of the theory.  Thus, a massive gravity theory
is distinct from Einstein gravity, even in the limit where the graviton
mass vanishes.  This discrepancy is a formulation of the
van~Dam--Veltman--Zakharov (VDVZ) discontinuity\cite{vDV,Z}.

The most accessible physical consequence of the VDVZ discontinuity is
the gravitational field of a star or other compact, spherically
symmetric source.  The ratio of the strength of the static (Newtonian)
potential to that of the gravitomagnetic potential is different for
Einstein gravity compared to massive gravity, even in the massless
limit.  Indeed the ratio is altered by a factor of order unity.  Thus,
such effects as light deflection by a star or perihelion precession of
an orbiting body would be affected significantly if the
graviton had even an infinitesimal mass.

This discrepancy appears for the gravitational field of any compact
object.  An even more dramatic example of the VDVZ discontinuity
occurs for a cosmic string.  A cosmic string has no static potential
in Einstein gravity; however, the same does not hold for a cosmic
string in massive tensor gravity.  One can see why using the momentum
space perturbative amplitudes for one-graviton exchange between two
sources $T_{\mu\nu}$ and $\tilde{T}_{\mu\nu}$:
\bea
	V_{massless}(q^2) &\sim& - {1 \over M_P^2}{1\over q^2}
	\left(T_{\mu\nu} - {1\over 2}\eta_{\mu\nu}T_\alpha^\alpha\right)
	\tilde{T}^{\mu\nu}	\\
	V_{massive}(q^2) &\sim& - {1 \over M_P^2}{1\over q^2+m^2}
	\left(T_{\mu\nu} - {1\over 3}\eta_{\mu\nu}T_\alpha^\alpha\right)
	\tilde{T}^{\mu\nu}
\eea
The potential between a cosmic string with
$T_{\mu\nu} = {\rm diag}(T,-T,0,0)$ and a test particle with
$\tilde{T}_{\mu\nu} = {\rm diag}(2\tilde{M}^2,0,0,0)$ is
\be
	V_{massless} = 0\ ,\ \ \ 
	V_{massive} \sim -{T\tilde{M}\over M_P^2}\ln r \ ,
\ee
where the last expression is taken in the limit $m \rightarrow
0$.  Thus in a massive gravity theory, we expect a cosmic string to
attract a static test particle, whereas in general relativity, no such
attraction occurs.  The attraction in the massive case can be
attributed to the exchange of the remnant light scalar mode that comes
from the decomposition of the massive graviton modes in the massless
limit.

Nevertheless, the presence of the VDVZ discontinuity is more subtle
than just described.  Vainshtein suggests that the discontinuity is
derived from only the lowest order, tree-level approximation and that
this discontinuity does not persist in the full classical theory
\cite{Vainshtein:1972sx}.  However, doubts remain
\cite{Boulware:1972my} since no self-consistent theory of massive
tensor gravity exists.  One can shed light on the issue of
nonperturbative continuity versus perturbative discontinuity by
studying a recent class of braneworld theories\footnote{ There has
been a recent revival of interest in the VDVZ discontinuity in the
context of braneworld theories.  These studies have focused on
variations of the Randall--Sundrum braneworld scenario where the
brane tension is slightly detuned from the bulk cosmological constant.
The localized four-dimensional graviton acquires a small mass,
allowing one to study the VDVZ problem in an effective massive
four-dimensional gravity theory.  For examples related to such work, see
\cite{Kogan:2000vb,Karch:2000ct,Porrati:2000cp,Karch:2001jb}.}
with a metastable graviton on the brane
\cite{Dvali:2000hr,Dvali:2001gm,Dvali:2001gx}.  The theory we wish
to consider has a four-dimensional brane embedded in a five-dimensional,
infinite volume Minkowski bulk, where the graviton is pinned to the
brane by intrinsic curvature terms induced by quantum fluctuations of
the matter.  The metastable
graviton has the same tensor structure as that for a massive graviton and
perturbatively has the same VDVZ problem in the limit that the
graviton linewidth vanishes.  In this model the momentum space
perturbative amplitude for one-graviton exchange is
\be
	V(q^2) \sim - {1 \over M_P^2}{1\over q^2+iqr_0^{-1}}
	\left(T_{\mu\nu} - {1\over 3}\eta_{\mu\nu}T_\alpha^\alpha\right)
	\tilde{T}^{\mu\nu}\ ,
\ee
where the scale $r_0$ is the scale over which the graviton evaporates off
the brane.  But unlike a massive gravity theory, this braneworld model
provides a self-consistent, generally covariant environment in which
to address the nonperturbative solutions in the limit as $r_0 \rightarrow
\infty$.  Indeed, exact cosmological solutions \cite{Deffayet} in this
theory already suggest that there is no VDVZ discontinuity at the
nonperturbative classical level \cite{Deffayet:2001uk}.

We would like to continue this program and investigate the
gravitational field of compact objects in the same braneworld theory
with a metastable brane graviton.  In this regard, one
would ideally like to identify the nonperturbative metric of a
spherical, Schwarzschild-like source.  That problem, however, possess
considerable, though not insuperable, computational difficulties.

Instead, we investigate the metric of a cosmic string as a close
alternative formulation of the VDVZ problem for a compact source.  The
advantage of this system is its relative simplicity, as well as the
clarity with which the VDVZ discontinuity manifests itself.  After
laying out the framework in which the problem is phrased, we identify
various regimes where one can linearize the cosmic string metric.  We
then argue that there exist a region where these cosmic string
solutions are simultaneously valid and that they are identical up to a
coordinate redefinition.  The resulting cosmic string metric indicates
there is no discontinuity in the fully nonperturbative theory.  It
also provides an understanding as to how different phases appear in
different regions near and far away from the string source.  We
conclude with some comments regarding the consequences of this
solution.

\section{The Solution}

\subsection{Preliminaries}

We wish to address the issues raised in the previous section using
a braneworld theory of gravity with an infinite volume bulk and a
metastable brane graviton
\cite{Dvali:2000hr}.  Consider a four-dimensional braneworld embedded
in a five-dimensional spacetime.  The bulk is empty; all energy-momentum
is isolated on the brane.  The action is
\be
S_{(5)} = -\frac{1}{2}M^3 \int d^5x \sqrt{|g|} \tilde{R} 
+\int d^4x \sqrt{-g^{(4)}}{\cal L}_m + S_{GH}\ .
\label{action}
\ee
The quantity $M$ is the fundamental five-dimensional Planck scale.
The first term in Eq.~(\ref{action}) corresponds to the
Einstein-Hilbert action in five dimensions for a five-dimensional
metric $g_{AB}$ (bulk metric) with Ricci scalar $R$.  The term
$S_{GH}$ is the Gibbons--Hawking action.  In addition, we consider an
intrinsic curvature term which is generally induced by radiative
corrections by the matter density on the brane \cite{Dvali:2000hr}:
\be
-\frac{1}{2}M^2_P \int d^4x \sqrt{-g^{(4)}}\ R^{(4)}\ .
\label{action2}
\ee
Here, $M_P$ is the observed four-dimensional Planck scale (see
\cite{Dvali:2000hr,Dvali:2001gm,Dvali:2001gx} for details).
Similarly, Eq.~(\ref{action2}) is the Einstein-Hilbert action for the
induced metric $g^{(4)}_{\mu\nu}$ on the brane, $R^{(4)}$ being its
scalar curvature.  The induced metric is
\be
g^{(4)}_{\mu\nu} = \partial_\mu X^A \partial_\nu X^B g_{AB}\ ,
\label{induced}
\ee
where $X^A(x^\mu)$ represents the coordinates of an event on the brane
labeled by $x^\mu$.

\begin{figure} \begin{center}\PSbox{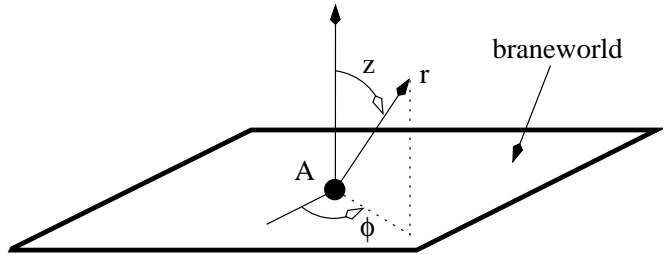
hscale=100 vscale=100 hoffset=-50 voffset=20}{2in}{1.5in}\end{center}
\caption{
A schematic representation of a spatial slice through a
cosmic string located at $A$.  The coordinate $x$ along the cosmic
string is suppressed.  The coordinate $r$ represents the 3-dimensional
distance from the cosmic string $A$, while the coordinate $z$ denotes
the polar angle from the vertical axis.  In the no-gravity limit,
the braneworld is the horizontal plane, $z = {\pi\over 2}$.  The
coordinate $\phi$ is the azimuthal coordinate.  Note that everywhere
except at the cosmic string, the unit vector in the direction of the
$z$-coordinate extends perpendicularly from the brane into the bulk.
}
\label{fig:flat}
\end{figure}

We wish to find the spacetime around a perfectly straight, infinitely
thin cosmic string.  With Lorentz boost symmetry and translational
invariance along the cosmic string, as well as rotational symmetry
around the string axis, the most general time-independent metric can
be written with the following line element:
\be
ds^2 = N^2(r,z)(dt^2-dx^2) - A^2(r,z)dr^2 - B^2(r,z)\left[dz^2
				+ \sin^2z\ d\phi^2 \right]\ ,
\label{metric}
\ee
where the string is located at $r = 0$ for all $(t,x)$.  These
coordinates are depicted in Fig.~\ref{fig:flat}.  If spacetime were
flat (i.e., $N=A=1$, $B=r$), we would choose the brane to be located
at $z = {\pi\over 2}$.  In general, one can choose coordinates within
the context of the line element Eq.~(\ref{metric}) such that the brane
is located at $z = {\pi\over 2}$, even when spacetime is not flat.
However, we will find it useful to apply a less stringent constraint,
considering coordinates where the brane is located at
$z ={\pi\alpha\over 2}$, where the parameter $\alpha$ is to be specified
by the brane boundary conditions.  Again, one can find a set of
coordinates within the ansatz Eq.~(\ref{metric}) in which this is
possible in general.

Assuming the cosmic string dominates the energy-momentum content of
the spacetime, we ignore the matter effects of the brane itself,
except through the intrinsic curvature term Eq.~(\ref{action2}).
Using the coordinate system specified by Eq.~(\ref{metric}), the
energy-momentum of the system is\footnote{Throughout this paper, we
define the distributional $\delta(x)$ of the variable $x$, such that
given any well-behaved function $f(x)$, $\int dx\ \delta(x)f(x) =
f(0)$\ .}
\be
	T_{tt} = -T_{xx} = N^2(r,z)
	{T\ \delta(r)\over 2\pi A(r,z)B^2(r,z)\sin z}\ ,
\label{matter}
\ee
where the parameter $T$ denotes the string tension and all other
components of the energy-momentum tensor are zero.
The Einstein equations dictated by the action
Eqs.~(\ref{action}--\ref{action2}) are
\be
	{1\over 2r_0}G_{AB}
	+ {1\over B(r,z)}
	\delta\left(z - {\pi\alpha\over 2}\right)G^{(4)}_{AB}
	= {1\over M_P^2}T_{AB}\ ,
\label{einstein}
\ee
where $G_{AB}$ is the five-dimensional Einstein tensor, $G^{(4)}_{AB}$
is the induced four-dimensional Einstein tensor on the brane,
$T_{AB}$ is the energy-momentum on the brane Eq.~(\ref{matter}),
and we have defined a crossover scale
\be
	r_0 = {M_P^2 \over 2M^3}\ .
\label{r0}
\ee
This scale characterizes that distance over which metric fluctuations
propagating on the brane dissipate into the bulk \cite{Dvali:2000hr}.

We assume a ${\cal Z}_2$--symmetric brane across $z = {\pi\alpha\over 2}$.
Under this circumstance, one may solve
Eqs.~(\ref{matter}--\ref{einstein}) by solving $G_{AB} = 0$ in the
bulk, i.e., when $z <{\pi\alpha\over 2}$ and $r\ne0$, such that the
following brane boundary conditions apply at $z = {\pi\alpha\over 2}$:
\bea
	\left({N_z\over N} + {A_z\over A} + {B_z\over B}\right)
	&=& {r_0r\over A^2}
	\left[{N_{rr}\over N} - {N_r\over N}{A_r\over A}
	+ {1\over r}\left({N_r\over N}-{A_r\over A}\right)\right]
	\nonumber	\\
	&\ &\ \ \ \ \ \ \ \ \ \ \ \ \ \ \ \ \ \
	\ \ \ \ \ \ \ \ -\  {\sqrt{1-\beta^2}\over\beta}
	\ +\ {r_0T/M_P^2\over 2\pi\beta}{1\over A}\delta(r)
	\nonumber	\\
	\left(2{N_z\over N} + {B_z\over B}\right)
	&=& {r_0r\over A^2}
	\left[{N^2_r\over N^2} + {2\over r}{N_r\over N}\right]
	\ -\  {\sqrt{1-\beta^2}\over\beta}
	\label{brane}							\\
	\left(2{N_z\over N} + {A_z\over A}\right)
	&=& {r_0r\over A^2}
	\left[2{N_{rr}\over N} + {N^2_r\over N^2}
	- 2{N_r\over N}{A_r\over A}\right]
	\nonumber
\eea
where we have defined $\beta = \sin{\pi\alpha\over 2}$ and where the
subscript represents partial differentiation with respect to the
corresponding coordinate.  Equations~(\ref{brane}) follow from
$G_{tt}$, $G_{rr}$ and $G_{\phi\phi}$, respectively, and are generated
by the intrinsic curvature term induced by the action
Eq.~(\ref{action2}).  We also impose boundary conditions to ensure
continuity of the metric and its derivatives at $z=0$, and to fix
a residual gauge degree of freedom by choosing $B(z={\pi\alpha\over 2}) = r$.

We wish to find the full five-dimensional spacetime metric induced by a
thin cosmic string situated within the braneworld.  The problem
defined by Eqs.~(\ref{matter}--\ref{einstein}) is dependent only on
the scale $r_0$ and the dimensionless parameter ${T\over M_P^2}$.  We
are interested in the problem when $r_0 \rightarrow \infty$ with all
other parameters held fixed.  Since $r_0$ represents the only scale in
the problem, this statement implies we are interested in the system when
$r \ll r_0$ with ${T\over M_P^2}$ fixed.

\subsection{The Einstein solution}

Before we attempt to solve the full five-dimensional problem given
by Eqs.~(\ref{matter}--\ref{einstein}) and Eq.~(\ref{brane}), it is
useful to review the cosmic string solution in simply four-dimensional
Einstein gravity \cite{Vilenkin:1981zs,Gregory:gh}.  For a cosmic
string with energy momentum Eq.~(\ref{matter}), the exact metric
may be represented by the line element:
\be
	ds^2 = dt^2 - dx^2
	- \left(1 - {T\over 2\pi M_P^2}\right)^{-2}dr^2 - r^2d\phi^2\ .
\label{einstein-soln}
\ee
This represents a flat space with a deficit angle ${T\over M_P^2}$.
Thus, there is no Newtonian potential between a cosmic string and a
static test particle.  However, a test particle (massive or massless)
suffers an azimuthal deflection of ${T\over M_P^2}$ when scattered
around the cosmic string.  With a different coordinate choice, the
line element can be rewritten as
\be
	ds^2 = dt^2 - dx^2
	- (y^2+z^2)^{-T/2\pi M_P^2}[dy^2 + dz^2]\ .
\ee
Again, there is no Newtonian potential between a cosmic string and a
static test particle.  However, in this coordinate choice, the
deflection of a moving test particle can be interpreted as resulting
from a gravitomagnetic force generated by the cosmic string.
We can ask whether this Einstein solution is recovered on the
brane in the limit of the theory where the graviton linewidth vanishes.
In this limit, gravity fluctuations originating on the brane are pinned
on that surface indefinitely, implying that gravity should resemble
a four-dimensional theory.  However, the question remains whether the
four-dimensional theory that results is Einstein gravity or some
massless scalar-tensor theory instead.

\subsection{Linearized Five-Dimensional Einstein Equations}

Let us examine the linearized form of the Einstein equations,
Eqs.~(\ref{einstein}).  We will see that the trick is to find an
appropriate background (including boundary conditions) around which
linearize.  We take the following:
\bea
	N(r,z) &=& 1 + n(r,z) + \cdots
	\nonumber \\
	A(r,z) &=& 1 + a(r,z) + \cdots
	\label{linearized} \\
	B(r,z) &=& r\left[1 + b(r,z)+ \cdots\ \right]
	\ ,	\nonumber
\eea
where it is assumed that the functions $\{n(r,z),a(r,z),b(r,z)\} \ll
1$ in the regimes of interest.  Taking the $R_{tt}$-component of the
Einstein equations, one find that the PDE for $n(r,z)$
\be
	r^2~n_{rr} + 2rn_r + n_{zz} + {\cos{z}\over\sin{z}}n_z = 0\ ,
\label{potential}
\ee
may be decoupled from the others.  Equation~(\ref{potential}) is
simply Laplace's equation for the Newtonian potential,
$n(r,z)$, reducing the determination of $g_{00}$ to a linear static
potential problem (albeit, with unusual boundary conditions
resulting from the presence of the brane).  In order
to determine $a(r,z)$ and $b(r,z)$, one can directly integrate the
$G_{zz}$ and the $G_{zz}$ components of the Einstein equations,
leaving
\bea
	a(r,z) &=& -2n(r,z) + rf'(r)\cos{z} + rg'(r) + g(r)
	\label{a}	\\
	b(r,z) &=& -2n(r,z) + f(r)\cos{z} + g(r)\ .
	\label{b}
\eea
The functions $f(r)$ and $g(r)$ are to be determined by the brane
boundary conditions, as well as the last remaining residual gauge.
This technique for decomposing the linearized Einstein equations is
the direct analog of that used in \cite{Gruzinov:2001hp}.

The brane boundary conditions Eqs.~(\ref{brane}) in the linearized
Einstein equations may be used to complete the determination of the
metric components.  The second two equations in Eqs.~(\ref{brane})
both yield
\be
	\beta f(r) = {\sqrt{1-\beta^2}\over\beta}
			- 2r_0n_r(r)|_{\sin z = \beta}\ .
\label{f}
\ee
Combining this equation for $f(r)$ and fixing the gauge choice
$b(r)|_{\sin{z} = \beta} = 0$ gives
\be
	g(r) = 2n(r)|_{\sin{z} = \beta}
		+ {\sqrt{1-\beta^2}\over\beta}r_0n_r(r)|_{\sin{z} = \beta}
		- {1-\beta^2\over\beta^2}\ .
\label{g}
\ee
The remaining equation in Eqs.~(\ref{brane}) is then used to
set the brane condition for the Newtonian potential $n(r,z)$:
\be
	n_z|_{\sin z =\beta} = r_0\left[\left(r 
			+ {2r_0\over 3}{\sqrt{1-\beta^2}\over\beta}
			\right)n_{rr}
		+n_r\right]_{\sin z =\beta}
	\ +\ {r_0\over 3\beta^2}\delta(r)
	\left[\beta + \left({T\over 2\pi M_P^2} - 1\right)\right]\ ,
\label{potential_brane}
\ee
where the coefficient of the $\delta$--function contribution to this
condition comes from the matter source (as reflected in the first
equation of Eqs.~(\ref{brane})) and from the step in $A(r,z)$ at
$r=0$ necessary to maintain elementary flatness at the location
of the string.  Once one determines $n(r,z)$ using Eq.~(\ref{potential}) and
the boundary conditions Eq.~(\ref{potential_brane}) as well as
$n_z(r)|_{z=0} = 0$, then one can automatically read off $a(r,z)$
and $b(r,z)$ using Eqs.~(\ref{a}--\ref{g}).

\subsection{The weak brane limit}

Let us first identify the solution to
Eqs.~(\ref{matter}--\ref{einstein}) and Eq.~(\ref{brane}) in the weak
field limit.  Here, we presume that the metric deviates from a flat
metric with a flat brane where the perturbations (of the bulk and the
brane) are proportional to the strength of the source, ${T\over M_P^2}$,
assuming this parameter is small.  With the coordinate
choice under consideration, one may keep the brane at $z = {\pi\over
2}$ while still allowing for a brane extrinsic curvature of ${\cal
O}({T\over M_P^2})$.  We refer to this limit where the extrinsic
curvature of the brane is perturbed around a $flat$ brane as the
weak brane limit.

In this limit one may use the linearized equations established in the
last subsection.  The explicit solution to Eqs.~(\ref{potential})
and~(\ref{potential_brane}) with $\beta = 1$ is
\be
	n(r,z) = -{r_0\over 3}\left({T\over 2\pi M_P^2}\right)
		\int_{0}^{\infty}
		{dk\over 1 + r_0k}~e^{-kr\cos z}J_0(kr\sin z)\ ,
\label{weakfield}
\ee
where $J_0$ is the usual Bessel function of the first kind.  One can
then solve for $a(r,z)$ and $b(r,z)$ directly using
Eqs.~(\ref{a}--\ref{g}).  One may also arrive at this result by
applying the graviton propagator \cite{Dvali:2000hr,Deffayet:2001uk}
and approximating the gravitational potential through one-particle
graviton exchange between the cosmic string source and a test particle
in a Minkowski spacetime with a flat braneworld.

Two limits are of interest.  The regime where $r \gg r_0$ represents
the crossover from four-dimensional to five-dimensional behavior
expected at the scale $r_0$.  Graviton modes localized on the brane
evaporate into the bulk over distances comparable to $r_0$.  The
presence of the brane becomes increasingly irrelevant as $r/r_0
\rightarrow \infty$ and a cosmic string on the brane acts as a
codimension-three object in the full bulk.  Here the metric is
asymptotically spherically symmetric (i.e., $z$-independent) while the
Newtonian potential Eq.~(\ref{weakfield}) becomes
\be
	n(r,z) = -\ {1\over 3}\left({T\over 2\pi M_P^2}\right){r_0\over r}
	\ +\ {\cal O}(r_0^2/r^2)\ .
\ee
Using the metric Eq.~(\ref{metric}), we find the metric on the brane
is specified by the line element 
\be
	ds^2 = N^2(r)|_{\sin z = 1}\ (dt^2-dx^2)
	- A^2(r)|_{\sin z = 1}\ dr^2 - r^2d\phi^2\ 
\label{metric-weak}
\ee
with
\bea
	N(r)|_{\sin z = 1} &=& 
	1\ -\ {1\over 3}\left({T\over 2\pi M_P^2}\right){r_0\over r}
	\ +\ {\cal O}(r_0^2/r^2)  \label{3d-weakN}\\
	A(r)|_{\sin z = 1} &=&
	1 \ +\ {2\over 3}\left({T\over 2\pi M_P^2}\right){r_0\over r}
	\ +\ {\cal O}(r_0^2/r^2)  \label{3d-weakA} \ ,
\eea
recovering the Schwarzschild-like solution for a codimension-three
object in five-dimensional spacetime with
\be
	r_G = {r_0T\over 2\pi M_P^2} = {T\over 4\pi M^3}
\ee
acting as the effective Schwarzschild radius.

In the complementary limit when $r \ll r_0$, we find that
Eq.~(\ref{weakfield}) becomes
\be
	n(r,z) = {1\over 3}\left({T\over 2\pi M_P^2}\right)
	\ln\left[{r\over r_0}(1 + \cos z)\right]
	\ +\ {\cal O}\left({T^2\over M_P^4}\right)\ .
\label{potential-weak}
\ee
Using the metric Eq.~(\ref{metric-weak}) on the brane, we find
\bea
	N(r)|_{\sin z = 1} &=& 1 
	\ +\ {1\over 3}\left({T\over 2\pi M_P^2}\ln{r\over r_0}\right)
	\ +\ {\cal O}\left({T^2\over M_P^4}\right)
	\label{2d-weakN} \\
	A(r)|_{\sin z = 1} &=& 1 \ +\ {2\over 3}{T\over 2\pi M_P^2}
	\ +\ {\cal O}\left({T^2\over M_P^4}\right)
	\label{2d-weakA}\ ,
\eea
which represents a conical space with deficit angle ${2\over 3}{T\over
M_P^2}$.  Recall that for pure four-dimensional Einstein gravity, this
metric is $N(r) = 1$ and $A(r) = (1-{T\over 2\pi M_P^2})^{-1}$, which
again represents a flat conical space with deficit angle ${T\over
M_P^2}$.  Thus in the weak brane limit, we expect not only an extra
light scalar field generating the Newtonian potential found in $N(r)$,
but also a discrepancy in the deficit angle with respect to the
Einstein solution.  Moreover, since
Eqs.~(\ref{2d-weakN}--\ref{2d-weakA}) deviate from the
four-dimensional Einstein solution, the brane boundary conditions
Eq.~(\ref{brane}) imply that {\em the brane itself possess an extrinsic
curvature whose magnitude is ${\cal O}({T\over M_P^2})$}.

We can ask the domain of validity of the solution
Eqs.~(\ref{2d-weakN}--\ref{2d-weakA}).  Examining the boundary conditions
Eqs.~(\ref{brane}) and the bulk Einstein equations $G_{AB} = 0$ and
comparing the size of terms neglected with respect to those included,
we see that on the brane this solution is only valid when
\be
	r_0\sqrt{{T\over M_P^2}} \ll r \ll r_0\ .
\label{limits-weak}
\ee
The left-hand inequality of Eq.~(\ref{limits-weak}) is the one of
interest.  For values of $r$ violating this condition, nonlinear
contributions to the Einstein tensor become important and the weak
brane approximation breaks down.  But this is precisely the regime we
are interested in, since we wish to understand what happens when $r$
and ${T\over M_P^2}$ are fixed and $r_0 \rightarrow \infty$.  We
need to find a solution in this regime.

\subsection{The $r/r_0 \rightarrow 0$ limit}

The weak brane approximation breaks down when the condition
Eq.~(\ref{limits-weak}) does not apply.  Outside this domain of
validity, nonlinear contributions to the Einstein equations become
important.  However, by relaxing the condition that the braneworld be
located at $\sin z = 1$, a perturbative solution to the Einstein
equations Eqs.~(\ref{matter}--\ref{einstein}) with the boundary
conditions Eq.~(\ref{brane}) can be found in the limit of interest
when $r \ll r_0$ with ${T\over M_P^2}$ held fixed.  We are still
interested in the limit of a weak source, i.e., ${T\over M_P^2}\ll 1$,
so that using the linearized equations establish in Sec.~IIC is still
applicable.

\begin{figure} \begin{center}\PSbox{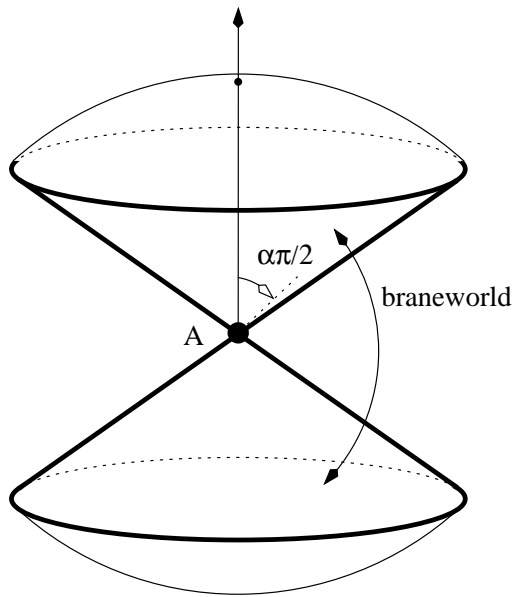
hscale=100 vscale=100 hoffset=-10 voffset=20}{2in}{3.25in}\end{center}
\caption{
A spatial slice through the cosmic string located at $A$.
As in Fig.~\ref{fig:flat} the coordinate $x$ along the cosmic string
is suppressed.  The solid angle wedge exterior to the cone is removed
from the space, and the upper and lower branches of the cone are
identified.  This conical surface is the braneworld ($z={\pi\alpha\over 2}$
or $\sin z = \beta$).  The bulk space now exhibits a deficit polar
angle (cf. Fig.~\ref{fig:flat}).  Note that this deficit in polar
angle translates into a conical deficit in the braneworld space.
}
\label{fig:space}
\end{figure}

Recall that the coordinate $z$ into the bulk acts as a polar angle,
but where the space it parametrizes has a deficit polar angle.  The
bulk is characterized by that part of space where $\sin z < \beta$
(i.e., $0 \le z < {\pi\alpha\over 2}$ and $\pi(1-{\alpha\over 2}) < z
\le \pi$) and the two surfaces where $\sin z = \beta$ are identified
and together represent the braneworld. The identification of these two
surfaces induces an extrinsic curvature contribution on the brane
which is compensated by the braneworld's intrinsic curvature.  Note
that the bulk is ${\cal Z}_2$--symmetric across the brane.  The key
difference between the analysis in this background and that in the
previous subsection is the inclusion of nonperturbative extrinsic
curvature in the background brane.  Figure~\ref{fig:space} depicts a
spatial slice through the cosmic string.

One begins by solving the linearized Newtonian potential problem
Eq.~(\ref{potential}) with the boundary conditions
Eq.~(\ref{potential_brane}) and $n_z(r)|_{z=0} = 0$.  Additional
boundary conditions are necessary.  In order to avoid the divergence
of the fields $\{n(r,z),a(r,z),b(r,z)\}$ near the origin that
leads to the importance of nonlinear contributions in the weak
brane limit, we choose
\be
	n(z)|_{r=0} = 0\ .
\label{bc1}
\ee
By specifying this boundary condition, one is required to constrain
$\beta$ (with the brane located at $\sin z = \beta$) for consistency
with the brane boundary condition, Eq.~(\ref{potential_brane}).  In
order to maintain consistency, the delta-function term must vanish,
requiring\footnote{
 The divergence in $n(r,z)$ as $r\rightarrow 0$ seen in the weak
 brane approximation is avoided because the matter source
 delta-function is matched by a step in the metric function $A(r,z)$
 rather than a logarithmic divergence in $N(r,z)$.}
\be
	\beta = 1 - {T\over 2\pi M_P^2} \ .
\label{alpha}
\ee
One last boundary condition needs to be specified for large-$r$.
We choose $n(r,z)$ to match asymptotically onto the form
\be
	n(z)|_{r=R} = {1\over 3}\left({T\over 2\pi M_P^2}\right)
	\ln\left[{R\over r_0}(1 + \cos z)\right]\ ,
\label{bc2}
\ee
for some $R$ such that $r_0\sqrt{T\over M_P^2}\ll R\ll r_0$.  We
choose this specific asymptotic form for reasons that will become
clear in the next section.  This boundary value problem for a static
solution to Laplace's equation Eq.~(\ref{potential}) is well-posed,
albeit in an unusual geometry and with unusual boundary conditions.

Unlike when $\sin z = 1$, this solution to Laplace's equation does not
possess a simple closed form.  However, one may articulate the dominant
contributions to the Newtonian potential in several key limits.  When
$r \ll r_0\sqrt{{T\over M_P^2}}$
\be
	n(r,z) = \left[{1\over 2}\cos z\right]{r\over r_0}
			- \left[{1\over 2}P_q(\cos z)\right]
				\left({r\over r_0}\right)^q + \cdots\ ,
\ee
with $q = 1+\sqrt{{T\over \pi M_P^2}}$ and $P_q(x)$ is the Legendre
function of the first kind of order $q$.  When
$r \gg r_0\sqrt{{T\over M_P^2}}$,
\be
	n(r,z) = {1\over 3}\left({T\over 2\pi M_P^2}\right)
	\ln\left[{3\over 2}{r\over r_0\sqrt{T\over 2\pi M_P^2}}
		(1 + \cos z)\right] + \cdots\ .
\ee
Indeed, one can arrive at an explicit form for the leading contribution
to the Newtonian potential on the brane itself:
\be
	n(r) = {1\over 3}\left({T\over 2\pi M_P^2}\right)
	\ln\left[1~+~{3\over 2}
	{r \over r_0\sqrt{T\over 2\pi M_P^2}}\right] + \cdots\ ,
\label{2d-potential}
\ee
when $r\ll r_0$.  One may ascertain $a(r,z)$ and $b(r,z)$ by directly
using Eqs.~(\ref{a}--\ref{g}).  By comparing the full Einstein
equations and brane boundary conditions,
Eqs.~(\ref{einstein}--\ref{brane}), with the linearized equations,
Eqs.~(\ref{linearized}--\ref{potential_brane}), one can confirm that
the Newtonian potential $n(r,z)$ is valid in the {\em entire} region
$r\ll r_0$, while the expressions for $a(r,z)$ and $b(r,z)$ using
Eqs.~(\ref{a}--\ref{g}) are valid when $r \ll r_0\sqrt{{T\over
M_P^2}}$ and when $r \gg r_0\sqrt{{T\over M_P^2}}$.

When $r \ll r_0\sqrt{T\over M_P^2}$, the metric on the brane is
determined by the line element
\be
	ds^2 = N^2(r)|_{\sin z = \beta}\ (dt^2-dx^2)
	- A^2(r)|_{\sin z = \beta}\ dr^2 - \beta^2r^2d\phi^2
\label{brane-metric}
\ee
with
\bea
        N(r)|_{\sin z =\beta} &=& 1 + {r\over 2r_0}\sqrt{T\over 2\pi M_P^2}
        \ +\  {\cal O}(r^2/r_0^2) \label{2d-braneN}      \\
        A(r)|_{\sin z =\beta} &=& 1 - {r\over 2r_0}\sqrt{T\over 2\pi M_P^2}
        \ +\  {\cal O}(r^2/r_0^2)\ , \label{2d-braneA}
\eea
where $\beta = 1 - \sqrt{T\over 2\pi M_P^2}$.  The brane solution,
Eqs.~(\ref{2d-braneN}--\ref{2d-braneA}), is distinct from the weak
brane solution, Eqs.~(\ref{2d-weakN}--\ref{2d-weakA}).  In particular,
the extrinsic curvature of the brane in the first case is
nonperturbative in the string tension, i.e., ${\cal O}(\sqrt{T\over
M_P^2})$, whereas extrinsic curvature of the brane in the weak brane
approximation is perturbative in the string tension, i.e., ${\cal
O}({T\over M_P^2})$.

The deficit polar angle in the bulk is $\pi(1-\alpha)$ where again
$\sin{\pi\alpha\over 2} = \beta$, while the deficit azimuthal angle in
the brane itself is $2\pi(1-\beta)$.  In the limit when $r_0
\rightarrow \infty$, the graviton linewidth vanishes and we recover a
flat conical space with a deficit angle $2\pi(1-\beta) = {T\over
M_P^2}$, the solution for a cosmic string in four-dimensional Einstein
gravity Eq.~(\ref{einstein-soln}).  Consequently, the cosmic string
solution Eqs.~(\ref{2d-braneN}--\ref{2d-braneA}) does not suffer from
a VDVZ discontinuity, supporting the results found for cosmological
solutions \cite{Deffayet:2001uk} in this braneworld theory with a
metastable brane graviton.

\section{Matching between Phases}

We wish to address the matching of the Einstein phase,
Eqs.~(\ref{2d-braneN}--\ref{2d-braneA}) to the weak brane phase,
Eqs.~(\ref{2d-weakN}--\ref{2d-weakA}).  These two
solutions are in distinct coordinate systems.  Nevertheless, when
the cosmic string source strength ${T\over M_P^2}$ is small, there
exists a region in $r$, namely $r_0\sqrt{T\over M_P^2} \ll r \ll r_0$,
where both solutions are valid.  In this section we show that there
exists a coordinate transformation that takes the Einstein phase
into the weak brane phase in this region, implying that these phases
are simply different parts of the same solution.

In the region $r_0\sqrt{T\over M_P^2} \ll r \ll r_0$, the Einstein
phase takes the form
\bea
	n(r,z) &=& {1\over 3}\left({T\over 2\pi M_P^2}\right)
	\ln\left[{3\over 2}
	{r\over r_0\sqrt{T\over 2\pi M_P^2}}(1 + \cos z)\right]
	+ \cdots	\nonumber	\\
	a(r,z) &=& {2\over 3}{r_0\over r}
		\left({T\over 2\pi M_P^2}\right)\cos z + \cdots
	\label{einstein-match}		\\
	b(r,z) &=& \left(\sqrt{T\over 2\pi M_P^2}
	- {2r_0\over 3r}{T\over 2\pi M_P^2}\right)\cos z + \cdots\ ,
			\nonumber
\eea
where the brane is located at $\sin z = 1 - {T\over 2\pi M_P^2}$ (or
$\cos z = \sqrt{T\over 2\pi M_P^2}$), and neglecting contributions of
${\cal O}({T \over M_P^2})$ but keeping leading order contributions,
e.g., $\sim {T\over 2\pi M_P^2}{r_0\over r}$.  The weak brane phase in
the same region has the form
\bea
	n(r,z) &=& {1\over 3}\left({T\over 2\pi M_P^2}\right)
	\ln\left[{r\over r_0}(1 + \cos z)\right]
	+ \cdots	\nonumber	\\
	a(r,z) &=& {2\over 3}{r_0\over r}
		\left({T\over 2\pi M_P^2}\right)\cos z + \cdots
	\label{weak-match}		\\
	b(r,z) &=& -{2\over 3}{r_0\over r}
		\left({T\over 2\pi M_P^2}\right)\cos z + \cdots\ ,
			\nonumber
\eea
where the brane is located at $\cos z = 0$, and where again
contributions of ${\cal O}({T \over M_P^2})$ are neglected.

We have chosen a region, $r_0\sqrt{T\over M_P^2} \ll r \ll r_0$,
such that the solution Eqs.~(\ref{einstein-match})
remains a valid solution to the linearized Einstein equations while
undergoing the following linear coordinate transformation.  Take a
new polar variable, ${\cal Z}$, such that
\be
	{\cal Z} = z + \sqrt{T\over 2\pi M_P^2}\sin z
			+ {\cal O}\left({T\over 2\pi M_P^2}\right)\ .
\label{z-trans}
\ee
Since
\be
	\sin{\cal Z} = \sin z\left[1 + \sqrt{T\over 2\pi M_P^2}\cos z\right]
			+ \cdots\ ,
\ee
the metric functions are still in the ansatz Eq.~(\ref{metric}) with the
new polar coordinate, ${\cal Z}$:
\be
	ds^2 = N^2(r,{\cal Z})(dt^2-dx^2) - A^2(r,{\cal Z})dr^2
			- B^2(r,{\cal Z})\left[d{\cal Z}^2
				+ \sin^2{\cal Z}\ d\phi^2 \right]\ .
\label{metric-new}
\ee
Under this coordinate redefinition, when the brane is located at
$\cos z = \sqrt{T\over 2\pi M_P^2}$, it is now located at
\be
	\cos{\cal Z} = \cos z - \sqrt{T\over 2\pi M_P^2} + \cdots
	 = 0 \ +\  {\cal O}\left({T\over 2\pi M_P^2}\right)\ .
\ee
and with a time renormalization
\be
	\tau = {t \over 1 - {1\over 6}\left({T\over 2\pi M_P^2}\right)
	\ln{T\over 2\pi M_P^2}}
\label{time}
\ee
the Einstein phase takes the form
\bea
	n(r,{\cal Z}) &=& {1\over 3}\left({T\over 2\pi M_P^2}\right)
	\ln\left[{r\over r_0}(1 + \cos{\cal Z})\right]
	+ \cdots	\nonumber	\\
	a(r,{\cal Z}) &=& {2\over 3}{r_0\over r}
		\left({T\over 2\pi M_P^2}\right)\cos{\cal Z} +\cdots	\\
	b(r,{\cal Z}) &=& -{2\over 3}{r_0\over r}
		\left({T\over 2\pi M_P^2}\right)\cos{\cal Z} + \cdots\ ,
			\nonumber
\eea
with the brane located at $\cos{\cal Z} = 0$, which is identical to
Eq.~(\ref{weak-match}) up to ${\cal O}({T \over M_P^2})$.  Thus, one
can match the Einstein phase solution
with the weak brane solution using the coordinate transformation
Eq.~(\ref{z-trans}) and the time redefinition Eq.~(\ref{time}),
implying that the Einstein phase in Sec.~IIE and the weak brane phase
in Sec.~IID are parts of the same solution.\footnote{On the brane
itself, the coordinate transformation is applicable {\em including}
${\cal O}({T\over 2\pi M_P^2})$.  One can confirm this by comparing
Eqs.~(\ref{2d-weakN}--\ref{2d-weakA}) with Eq.~(\ref{2d-potential})
and Eqs.~(\ref{a}),~(\ref{f}), and~(\ref{g}) using the time renormalization
Eq.~(\ref{time}).}

\section{Discussion}

In different parametric regimes, we find different qualitative
behaviors for the brane metric around a cosmic string.  For an
observer at a distance $r \gg r_0$ from the cosmic string, where
$r_0^{-1}$ characterizes the graviton's effective linewidth, the
cosmic string appears as a codimension-three object in the full bulk.
The metric is Schwarzschild-like in this regime.  When $r \ll r_0$,
brane effects become important, and the cosmic string appears as a
codimension-two object on the brane.  When the source is weak (i.e.,
when the tension, $T$, of the string is much smaller than the square
of the four-dimensional Planck scale, $M_P^2$), the Einstein solution
with a deficit angle of ${T\over M_P^2}$ holds on the brane when
$r \ll r_0\sqrt{{T\over M_P^2}}$.  In the region on the brane when $r
\gg r_0\sqrt{{T\over M_P^2}}$ (but still where $r \ll r_0$), the weak
brane approximation prevails, the cosmic string exhibits a
nonvanishing Newtonian potential and space suffers a deficit angle
different from ${T\over M_P^2}$.

We identified a coordinate transformation connecting the weak brane
phase, Eqs.~(\ref{2d-weakN}--\ref{2d-weakA}), and the Einstein phase,
Eqs.~(\ref{2d-braneN}--\ref{2d-braneA}).  Each phase is a linear
solution which becomes strongly nonlinear outside of its domain of
validity simply because the corresponding coordinate system in which
each solution is linear differs from the other.  The full nonlinear
solution in this light is reminiscent of the ansatz introduced by
Vainshtein \cite{Vainshtein:1972sx} for the Schwarzschild solution in
a massive gravity theory.  Moreover, the presence of this weak brane
phase at large distances from the cosmic string may have
non-negligible consequences for the observational search for cosmic
strings through gravitational lensing techniques \cite{lensing}.  For
a GUT scale ($10^{16}$~GeV) cosmic string, the Einstein deficit angle
for the string is $\sim 10^{-5}$.  This implies that the light
deflection by the string differs significantly from the predictions of
general relativity at distances
\be
	r \sim r_0\sqrt{{M_{\rm GUT}^2\over M_P^2}}
	\sim 3\ {\rm Mpc}\ ,
\ee
where $r_0$ is taken to be today's Hubble radius.  Such a seemingly
peculiar choice of $r_0$ is intriguing cosmologically
\cite{Deffayet,Deffayet:2001pu}.  Detailed discussion on how such a
correspondingly small fundamental Planck scale is
possible without serious phenomenological obstacles may be found in
\cite{Dvali:2001gm,Dvali:2001gx}.

The solution presented here supports the Einstein solution near the
cosmic string in the limit that $r_0 \rightarrow \infty$.  This
observation suggests that the braneworld theory under consideration
does not suffer from a van~Dam--Veltman--Zakharov (VDVZ)
discontinuity, corroborating the findings for cosmological solutions
in the same theory \cite{Deffayet:2001uk}.  Far from the source, the
gravitational field is weak, and the geometry of the brane within the
bulk is not substantially altered by the presence of the cosmic
string.  Propagation of the light scalar mode is permitted.  However
near the source, the gravitational fields induce a nonperturbative
extrinsic curvature in the brane.  That extrinsic curvature suppresses
the coupling of the scalar mode to matter and only the tensor mode
remains, thus Einstein gravity is recovered.  As one takes
$r_0 \rightarrow \infty$, the region where the source induces a large
brane extrinsic curvature grows with $r_0$, implying Einstein gravity
is strictly recovered in this limit.

In this paper, we investigate the spacetime around a cosmic string on
a brane in a five-dimensional braneworld theory that supports a
metastable brane graviton.  This system has the advantage
of offering a semianalytic solution to the metric around a compact
object, while still providing a clear example in which the VDVZ
discontinuity manifests itself.  The result may help shed light on the
more difficult, more immediately relevant problem of a
Schwarzschild-like spherical source in this braneworld theory.  At the
same time, the cosmic string solution is itself interesting and,
should these objects exist in nature, would have testable
phenomenological consequences.

\acknowledgments

The author would like to thank C.~Deffayet, G.~Dvali, A.~Gruzinov,
M.~Porrati, R. Scoccimarro and E.~J.~Weinberg for helpful discussions.
This work is sponsored in part by NSF Award PHY-9996137 and the
David and Lucille Packard Foundation Fellowship 99-1462.

\end{document}